\documentclass[english]{tMOP2e}
\usepackage[T1]{fontenc}

\usepackage{graphicx}

\usepackage{epstopdf}
\theoremstyle{plain}

\theoremstyle{definition}

\theoremstyle{remark}

\usepackage{babel}
\begin{document}

\title{Multimode Raman light-atom interface in warm atomic ensemble as
multiple three-mode quantum operations}

\author{\name{Micha\l{} Parniak\textsuperscript{a}$^{\ast}$\thanks{$^\ast$Corresponding author. Email: michal.parniak@fuw.edu.pl}, Daniel P\k{e}cak\textsuperscript{b} and Wojciech Wasilewski\textsuperscript{a}}
\affil{\textsuperscript{a}Institute of Experimental Physics, Faculty of Physics, University
of Warsaw, Pasteura 5, 02-093 Warsaw, Poland;
\textsuperscript{b}Institute of Physics, Polish Academy of Sciences, Al. Lotników 32/46,
02-668 Warsaw, Poland}
}






\maketitle

\begin{abstract}
We analyze the properties of a Raman quantum light-atom interface
in long atomic ensemble and its applications as a quantum memory or
two-mode squeezed state generator. We consider the weak-coupling regime and include both Stokes and anti-Stokes
scattering and the effects of Doppler broadening in buffer gas assuming
frequent velocity-averaging collisions. We find the Green functions
describing multimode transformation from input to output fields of
photons and atomic excitations. Proper mode basis is found via singular
value decomposition for short interaction times. It reveals that triples of modes are coupled
by a transformation equivalent to a combination of two beamsplitters
and a two-mode squeezing operation. We analyze the possible transformations
on an example of warm rubidium-87 vapor. 
The model we present bridges
the gap between the Stokes only and anti-Stokes only interactions
providing simple, universal description in a temporally and longitudinally
multimode situation. Our results also provide an easy way to find
an evolution of the states in a Schrödinger picture thus facilitating
understanding and design. 
\end{abstract}


\section{Introduction}

The off-resonant Raman interface is a vividly developing approach
to quantum memory \cite{Reim2011,Heshami2014}. An off-resonant anti-Stokes
scattering in optically thick atomic vapor can be used to transfer
excitations from the optical field to atoms in the read-in process
\cite{Michelberger2015}. Atomic coherence that takes on the form
of spinwave is created and can be converted back to light in the read-out
process \cite{VanderWal2003,Bashkansky2012}. The spinwaves can be
stored and then further manipulated \cite{Hosseini2009}. A similar
approach incorporates cold atomic ensembles instead of warm vapors
\cite{Bimbard2014,Ding2015a}. The Raman process is perhaps the simplest
out of many possible realizations of quantum memories. In addition,
a Stokes transition can be driven by properly tuning the pump beam,
which leads to the spontaneous creation of excitations pairs in the
write-in process, where for each photon a single atom is excited in
a spinwave \cite{Duan2001}. The interface also lends itself to a
multispatial mode use \cite{Chrapkiewicz2012,Koodynski2012,Higginbottom2012}.
Simultaneous driving of both Stokes and anti-Stokes transitions
leads to a number of interesting phenomena enabling quantum engineering \cite{Dabrowski2014,DeEchaniz2008,Brannan2014,Wu2010}. When
they are in perfect balance, quantum non-demolition (QND) measurement
type of interaction occurs \cite{Wasilewski2009,Sewell2013}. 

Here we focus on the effects of coexistence of both Stokes and anti-Stokes
scattering in the presence of Doppler broadening. We analyze the situation
for a wide range of detunings resulting in rather arbitrary proportions
between elementary processes. It turns out that in the off-resonant,
high-optical-depth case when losses can be neglected the combined
action of the Stokes and anti-Stokes scattering creates similar spatiotemporal
mode structure as each of them separately would \cite{Wasilewski2006,Raymer2004}.
However, the transformation of the annihilation/creation operators
of respective modes is more involved. It can be decomposed into a
pair of beamsplitters and a squeezing operation between pairs of modes:
atomic, photonic Stokes and photonic anti-Stokes. In particular we
show that both read-out from atomic memory to light and read-in of
light state into the memory typically involves significant squeezing
of the quantum state. The squeezing can be only avoided when anti-Stokes
coupling is much higher than the Stokes. The model we present facilitates understanding and design under realistic experimental
conditions.

This paper is organized as follows. In Section \ref{sec:Evolution-of-fields}
we write out equations of evolution of coupled quantum fields. We
include effects of frequent collisions with buffer gas. The solution
is given in a form of the input-output relations for which mode basis
diagonalizing the relations exists.
We analyze the quantum evolution of the coupled triples of modes.
In Section \ref{sec:Decomposition-of-three} we rewrite the interaction
of three fundamental modes in a form easiest to comprehend and analyze.
Section \ref{sec:Conclusions} concludes the paper.

\section{Evolution of fields\label{sec:Evolution-of-fields}}

\subsection{Maxwell-Bloch equations}

\begin{figure}
\centering{}\includegraphics[scale=0.8]{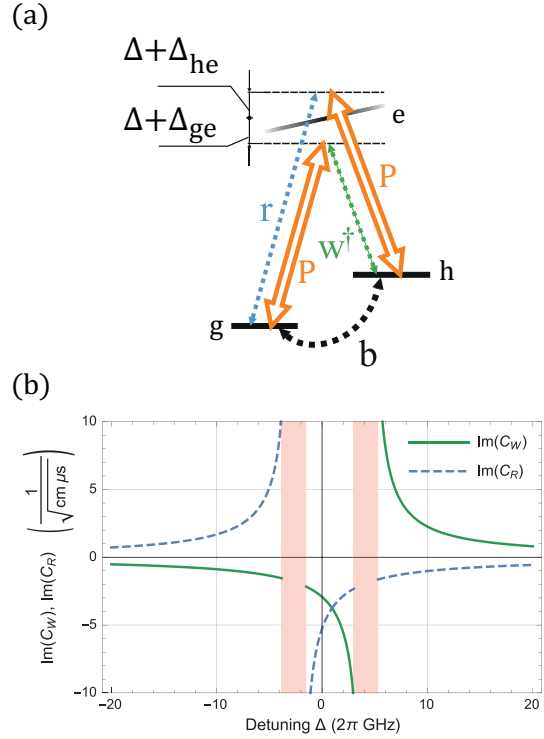}\protect\caption{(color online) (a) Off-resonant Raman scattering in a 3-level atom.
Linearly polarized pump of amplitude $P$ couples two separate
hyperfine components of the ground state, $|g\rangle$ and $|h\rangle$
via the excited state $|e\rangle$. Pump is detuned from the $|g\rangle-|e\rangle$
transition by $\Delta+\Delta_{ge}$ and from the $|h\rangle-|e\rangle$
transition by $\Delta+\Delta_{he}$, where $\Delta=0$ means the pump
is tuned to the line centroid. Tilting and shading of the upper level
symbolizes the Doppler shift and atomic population distribution respectively.
(b) Imaginary parts of coherent scattering coefficients $c_{R}(\Delta)$,
$c_{W}(\Delta)$ for rubidium 87 D1 line ($\lambda\approx795$~nm)
as a function of detuning $\Delta$ between the driving field and
the centroid of the line. We take two hyperfine components of the
ground state as levels $|g\rangle=|F=1,m_{F}=0\rangle$ and $|h\rangle=|F=2,m_{F}=0\rangle$.
We take the atom number density equal $N=10^{12}$~cm$^{-3}$ and
pump field Rabi frequency equal to the natural linewidth $\Gamma/2\pi=5.75$
MHz. Real parts of the coherent scattering coefficients are found
to be much smaller than the plotted imaginary parts and consequently
we exclude them from consideration. Shaded regions close to resonances
(closer than three times the Doppler width) where losses and pump
absorption have significant effect are excluded from the analysis.
\label{fig:3level}}
\end{figure}

We consider a spatially extended ensemble of length $L$ of three
or more level atoms moving with different velocities but otherwise
identical. The atoms are assumed to have two long-lived ground levels
$|g\rangle$ and $|h\rangle$ used to store quantum information. The
Raman interaction is driven by intense classical field $P$ in a square-shaped
pulse of duration $T$. We consider only forward scattering and thus
assume no dependence of the fields on perpendicular coordinates. We
also neglect all kinds of decoherence which is justified for large
optical depths. Initially all atoms are assumed to occupy $|g\rangle$
level. As illustrated in Fig.~\ref{fig:3level}(a) the interaction
couples the atoms to weak quantum fields (sidebands) $\hat{{r}}(z,t)$
and $\hat{{w}}^{\dagger}(z,t)$, corresponding to anti-Stokes and
Stokes modes, respectively. 

The beams are assumed to be virtually collinear and therefore only
the velocity $v$ of the atoms along the beams affects the interaction.
The splitting of the ground levels is taken to be so small that the
two-photon resonance condition is simultaneously met for all velocity
classes. However the detuning of the pump $P$ field
from resonance is different for each velocity class and therefore
the evolution of the atomic state is somewhat different. Thus we take
the atomic coherence field operator $\hat{{b}}(z,t,v)$, understood
within the Holstein-Primakoff approximation \cite{Holstein1940},
to be velocity class dependent\footnote{The atomic field operator is normalized so that the commutator $[\hat{{b}}(z,t,v),\hat{{b}}^{\dagger}(z',t,v')]=\delta(t-t')\delta(v-v')$,
therefore $\hat{{b}}(z,t,v)=\frac{{1}}{\sqrt{{N\Delta z\Delta v}}}\sum_{i}|g_{i}\rangle\langle h_{i}|$,
where the summation runs through $N$ atoms in slice $\Delta z$ and velocity
class $\Delta v$.}.

Together with warm atomic ensembles an inert buffer gas, such as neon,
krypton or xenon is used to make the motion of atoms diffusive and
consequently prolong the lifetime of stored spatial mode \cite{Parniak2014,Chrapkiewicz2014b}.
The presence of such buffer gas inherently leads to multiple effects
affecting quantum memories \cite{Manz2007}, one of which are the
velocity changing collisions. These collisions transfer atoms from
velocity class $v'$ into $v$ with probability per unit time given
by $\gamma_{v}K(v\leftarrow v')\mathrm{{d}}v'$, where $\gamma_{v}$
is the collision rate and $K(v\leftarrow v')$ is the collisional
kernel \cite{McGuyer2012}. 

Under the assumption that the driving field $P$ is off-resonant with
given detuning from line centroid $\Delta$, the excited level can
be adiabatically eliminated. The interaction between the atoms and
the sidebands is given by a set of Maxwell-Bloch equations \cite{Raymer1981}.
They can be cast in terms of field operators in a reference frame
co-moving with weak light \cite{Raymer2004}:

\begin{equation}
\frac{\partial\hat{{r}}(z,t)}{\partial z}=\int\sqrt{g(v)}c_{R}(\Delta+kv)\hat{{b}}(z,t,v)\mathrm{{d}}v\label{eq:r}
\end{equation}

\begin{equation}
\frac{\partial\hat{w}^{\dagger}(z,t)}{\partial z}=\int\sqrt{g(v)}c_{W}(\Delta+kv)\hat{{b}}(z,t,v)\mathrm{{d}}v\label{eq:w}
\end{equation}

\begin{equation}
\begin{aligned}\frac{\partial\hat{{b}}(z,t,v)}{\partial t}=\sqrt{g(v)}c_{W}^{*}(\Delta+kv)\hat{{w}}^{\dagger}(z,t)\\
+\sqrt{g(v)}c_{R}(\Delta+kv)\hat{{r}}(z,t)-s(\Delta+kv)\hat{{b}}(z,t,v)\\
+\gamma_{v}\int K(v\leftarrow v')\sqrt{{\frac{{g(v')}}{g(v)}}}\hat{{b}}(z,t,v')\mathrm{{d}}v',
\end{aligned}
\label{eq:b}
\end{equation}

where $g(v)=\frac{{1}}{\sqrt{{2\pi\sigma^{2}}}}\exp({-\frac{v^{2}}{2\sigma^{2}}})$
is the thermal velocity distribution
and $k=\frac{{2\pi}}{\lambda}$ is the light wavevector. The interaction
is parametrized by coherent scattering coefficients $c_{R}(\Delta)$
and $c_{W}(\Delta)$ plotted in Fig.~\ref{fig:3level}(b) and given
by:

\begin{equation}
c_{R}(\Delta)=-\sqrt{{\frac{{N\omega}}{2\hbar c\epsilon_{0}}}}\sum_{e}\frac{{Pd_{ge}d_{eh}}}{(\Gamma-2i(\Delta+\Delta_{he}))\hbar}\label{eq:cr}
\end{equation}

\begin{equation}
c_{W}(\Delta)=-\sqrt{{\frac{{N\omega}}{2\hbar c\epsilon_{0}}}}\sum_{e}\frac{{Pd_{he}d_{eg}}}{(\Gamma+2i(\Delta+\Delta_{ge}))\hbar}\label{eq:cw}
\end{equation}

where $\omega=\frac{{2\pi c}}{\lambda}$, $\Delta_{ij}$ is the difference
between resonant frequency of $|i\rangle-|j\rangle$ transition and
line central frequency (also see Fig.~\ref{fig:3level}(a) for reference),
$d_{ij}$ is the dipole moment for this transition and physical constants
are denoted as usual. The sum runs through all possible excited states $|e\rangle$. Limiting cases correspond to $|c_{W}|\ll|c_{R}|$
for anti-Stokes scattering and $|c_{R}|\ll|c_{W}|$ for Stokes scattering.
If the pump is far-detuned, so that $\Delta+\Delta_{ij}\gg\Gamma$,
we find the imaginary parts of coherent scattering coefficients $c_{R}$
and $c_{W}$ to be strongly dominant over the real parts, under the
assumption that pump amplitude $P$ is real. Indeed, the classical
field amplitude $P$ can be always made real by suitable shift of
the time reference. The same assumptions hold for the differential Stokes shift $s(\Delta)$ given by:

\begin{equation}
\begin{aligned}
s(\Delta)=\sum_{e}\frac{{|Pd_{eh}|^{2}}}{2(\Gamma-2i(\Delta+\Delta_{he}))\hbar^{2}}\\
+\frac{{|Pd_{eg}|^{2}}}{2\left(\Gamma+2i(\Delta+\Delta_{ge})\right)\hbar^{2}}.\label{eq:s}
\end{aligned}
\end{equation}

\subsection{Fast collisions approximation}

The inclusion of velocity changing collision kernel in Eq.~(\ref{eq:b})
may severely influence the solution, as different Maxwellian velocity
distribution preserving kernels might be used \cite{McGuyer2012,Marsland2012,Kryszewski1997}.
However, if we assume that the collisions are fast compared to the
Raman interaction $\gamma_{v}\gg Lc_{R,W}^{2}$, then the velocity
dependence of number operator of atoms in $|h\rangle$ state represented
by $\hat{{b}}^{\dagger}(z,t,v)\hat{{b}}(z,t,v)$ remains close to
thermal equilibrium. Consequently, we may separate out the known,
Gaussian velocity dependence and assume $\hat{{b}}(z,t,v)=\hat{{b}}(z,t)\sqrt{{g(v)}}$.

The equation~(\ref{eq:b}) for $\hat{{b}}(z,t,v)$ can now be integrated
formally and averaged over velocity distribution $\hat{{b}}(z,t)=\int\sqrt{{g(v)}}\hat{{b}}(z,t,v)\mathrm{{d}}v$,
yielding:

\begin{equation}
\begin{aligned}
\hat{{b}}(z,t)=\hat{{b}}(z,0)\\
+\int_{0}^{t}e^{-(t-t')\bar{{s}}}\left(\bar{{c}}_{R}\hat{{r}}(z,t')+\bar{{c}}_{W}^{*}\hat{{w}}^{\dagger}(z,t')\right){\mathrm{{d}}}t',\label{eq:b-sol1}
\end{aligned}
\end{equation}

where the differential Stokes shift $s(\Delta)$ and coherent scattering
coefficients $c_{R}(\Delta)$ and $c_{W}(\Delta)$ are now averaged
according to the formula $\bar{{x}}=\int x(\Delta+kv)g(v)\mathrm{{d}}v$.
Substituting this result to Eqs.~(\ref{eq:r})--(\ref{eq:w}) gives
the following equation of evolution for the photonic modes:

\begin{equation}
\begin{aligned}\frac{\partial}{\partial z}\begin{pmatrix}\hat{r}(z,t)\\
\hat{w}^{\dagger}(z,t)
\end{pmatrix}=\begin{pmatrix}\overline{c}_{\mathrm{R}}\\
\overline{c}_{\mathrm{W}}
\end{pmatrix}\hat{b}(z,0)e^{-\overline{s}t}\\
+\int_{0}^{t}e^{-(t-t')\overline{s}}\begin{pmatrix}\bar{{c}}_{R}^{2} & \bar{{c}}_{R}\bar{{c}}_{W}^{*}\\
\bar{{c}}_{W}\bar{{c}}_{R} & |\bar{{c}}_{W}|^{2}
\end{pmatrix}\begin{pmatrix}\hat{r}(z,t')\\
\hat{w}^{\dagger}(z,t')
\end{pmatrix}\mathrm{{d}}t'.
\end{aligned}
\label{eq:rw-sol1}
\end{equation}

Note the average atomic coefficients $\bar{c}_{W}$, $\bar{c}_{R}$
and $\bar{s}$ are virtually purely imaginary for far-detuned pump
field. From now on we will assume these coefficients are imaginary.

\subsection{Input-output relations}

The solution of equations~(\ref{eq:b-sol1})--(\ref{eq:rw-sol1})
takes on a form of a linear transformation between the
input and output quantum fields. First, a squeezing transformation
i.e. hyperbolic rotation $R(\zeta)$ between the photonic modes is
used to diagonalize the matrix from Eq.~(\ref{eq:rw-sol1}). Modes
$\hat{{c}}^\dagger(z,t)$ and $\hat{{d}}(z,t)$ are defined as linear combinations
of $\hat{{r}}(z,t)$ and $\hat{{w}}^{\dagger}(z,t)$ according to
the formulas:

\begin{equation}
\begin{pmatrix}\hat{r}(L,t)\\
\hat{w}^{\dagger}(L,t)
\end{pmatrix}=R^{-1}(\zeta)\begin{pmatrix}\hat{c}^{\dagger}(L,t)\\
\hat{d}(L,t)
\end{pmatrix}\label{eq:rw_out=00003DS_cd}
\end{equation}

\begin{equation}
\begin{pmatrix}\hat{c}^{\dagger}(0,t)\\
\hat{d}(0,t)
\end{pmatrix}=R(\zeta)\begin{pmatrix}\hat{r}(0,t)\\
\hat{w}^{\dagger}(0,t)
\end{pmatrix}\label{eq:rw_out=00003DS_cd-1}
\end{equation}

where the diagonalizing hyperbolic rotation matrix is defined as

\begin{equation}
R(\zeta)=\begin{cases}
\begin{pmatrix}\mathrm{{sinh}}(\zeta) & \mathrm{{cosh}}(\zeta)\\
\mathrm{{cosh}}(\zeta) & \mathrm{{sinh}}(\zeta)
\end{pmatrix} & \begin{matrix}\mathrm{{with\:}}\zeta=\mathrm{{atanh}}|\frac{{\bar{{c}}_{R}}}{\bar{{c}}_{W}}| \\ \mathrm{{for}\:}|\bar{{c}}_{W}|>|\bar{{c}}_{R}|\end{matrix}\\
\\
\begin{pmatrix}\mathrm{{cosh}}(\zeta) & \mathrm{{sinh}}(\zeta)\\
\mathrm{{sinh}}(\zeta) & \mathrm{{cosh}}(\zeta)
\end{pmatrix} & \begin{matrix}\mathrm{{with\:}}\zeta=\mathrm{{atanh}}|\frac{{\bar{{c}}_{W}}}{\bar{{c}}_{R}}| \\ \mathrm{{for}\:}|\bar{{c}}_{W}|<|\bar{{c}}_{R}|\end{matrix}
\end{cases}\label{eq:s1-dzeta-1}
\end{equation}

Note a significant difference between the two cases from above equation.
When $|\bar{{c}}_{W}|>|\bar{{c}}_{R}|$ both $\hat{c}(z,t)$ and $\hat{d}(z,t)$
are annihilation operators. On the contrary when $|\bar{c}_{W}|<|\bar{{c}}_{R}|$
both $\hat{c}(z,t)$ and $\hat{d}(z,t)$ play role of bosonic creation
operators. We can generally write $[\hat{c}(z,t),\hat{c}^{\dagger}(z,t')]=[\hat{d}(z,t),\hat{d}^{\dagger}(z,t')]=\mathrm{{sgn}}(|\bar{{c}}_{W}|-|\bar{{c}}_{R}|)\delta(t-t')$.
In particular in the limiting cases of pure Stokes or pure anti-Stokes
scattering corresponding to $|\bar{c}_{W}|\gg|\bar{{c}}_{R}|$ and
$|\bar{c}_{W}|\ll|\bar{{c}}_{R}|$, the $R(\zeta)$ is an identity
or mode-swap transformation, respectively.

The mode $\hat{{d}}(z,t)$ turns out to be decoupled and constant,
thus $\hat{{d}}(L,t)=\hat{{d}}(0,t)$. The equations coupling atoms
$\hat{{b}}(z,t)$ with photonic mode $\hat{{c}}(z,t)$ are the same
as in single-sideband Raman scattering \cite{Raymer1981,Raymer1985a,Raymer2004}
and their solution reads:
\begin{equation}
\begin{aligned}
\hat{b}(z,T)=\int_{0}^{T}\hat{{c}}^{\dagger}(0,t')\sqrt{\kappa}\Sigma_{0}(z,T-t')\mathrm{{d}}t'\\
+\int_{0}^{z}\hat{{b}}(z',0)[e^{-T\bar{{s}}}\delta(z-z')+T\Sigma_{1}(z-z',T)]\mathrm{{d}}z'
\label{eq:b-sol2-1}
\end{aligned}
\end{equation}

\begin{equation}
\begin{aligned}
\hat{c}^{\dagger}(L,t)=\int_{0}^{L}\hat{{b}}(z',0)\sqrt{\kappa}\Sigma_{0}(L-z',t)\mathrm{{d}}z'\\
+\int_{0}^{t}\hat{{c}}^{\dagger}(0,t')[L\Sigma_{1}(L,t-t')+\delta(t-t')]\mathrm{{d}}t'\label{eq:c-sol2-1}
\end{aligned}
\end{equation}

where the interaction strength is measured by

\begin{equation}
\kappa=\overline{c}_{\mathrm{R}}^{2}+|\overline{c}_{\mathrm{W}}|^{2}\label{eq:kappa-def}
\end{equation}

and $\Sigma_{1}(z,t)=\sqrt{{\frac{{\kappa}}{zt}}}I_{1}(2\sqrt{{\kappa zt}})$,
$\Sigma_{0}(z,t)=I_{0}(2\sqrt{{\kappa zt}})$, where $I_{n}(x)$ are
the modified Bessel functions of the first kind.

\subsection{Singular modes\label{sub:Singular-modes}}

\begin{figure}
\begin{centering}
\includegraphics[scale=1]{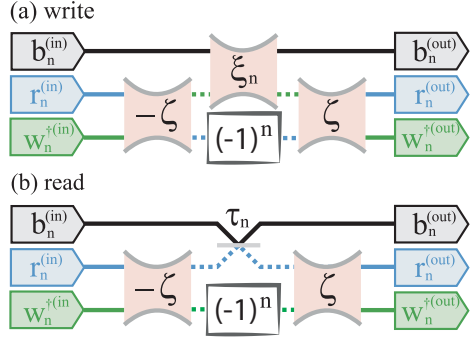}\protect\caption{(color online) Decompositions of interaction between $n$-th coupled
triple of modes into hyperbolic rotations $R(\zeta)$ and squeezing
by $\exp(\xi_{n})$ or beamsplitter with transmission $\tau_{n}$.
(a) Stokes domination $|\bar{{c}}_{W}|>|\bar{{c}}_{R}|$ {[}corresponding
matrix decomposition given by Eq.~(\ref{eq:g-sq}){]} and (b) anti-Stokes
domination $|\bar{{c}}_{W}|<|\bar{{c}}_{R}|$ {[}corresponding matrix
decomposition given by Eq.~(\ref{eq:g-bs}){]}.\label{fig:rw-decompos}}
\end{centering}
\end{figure}

The input-output relations in Eqs.~(\ref{eq:b-sol2-1}) and (\ref{eq:c-sol2-1})
can be simplified by suitable choice of mode basis for atomic $u_{n}^{({\rm in/out)}}(z)$
and photonic $v_{n}^{({\rm in/out)}}(t)$ fields in which $n$-th
atomic mode mixes only with photonic modes with the same number \cite{Raymer2004,Wasilewski2006,Koodynski2012}.
The annihilation operators for modes before and after the interaction
are defined as follows: $\hat{b}_{n}^{\mathrm{(in)}}=\int u_{n}^{(\mathrm{in})}(z)\hat{b}(z,0)\mathrm{{d}}z$,
$\hat{r}_{n}^{\mathrm{(in)}}=\int v_{n}^{(\mathrm{in})}(t)\hat{r}(0,t)\mathrm{{d}}t$,
$\hat{w}_{n}^{\dagger\mathrm{(in)}}=\int v_{n}^{(\mathrm{in})}(t)\hat{w}^{\dagger}(0,t)\mathrm{{d}}t$,
$\hat{b}_{n}^{\mathrm{(out)}}=\int u_{n}^{(\mathrm{out})}(z)\hat{b}(z,T)\mathrm{{d}}z$,
$\hat{r}_{n}^{\mathrm{(out)}}=\int v_{n}^{(\mathrm{out})}(t)\hat{r}(L,t)\mathrm{{d}}t$
and $\hat{w}_{n}^{\dagger\mathrm{(out)}}=\int v_{n}^{(\mathrm{out})}(t)\hat{w}^{\dagger}(L,t)\mathrm{{d}}t$.
The resulting operator, e.g. $\hat{{b}}_{n}^{(\mathrm{{in})}}$, annihilates
a single excitation in the respectful mode, in this case spatial mode
$u_{n}^{(\mathrm{{in}})}(z)$.  The modes are orthonormal and the commutator $[b_m^{(\mathrm{in})}, b_n^{(\mathrm{in})\dagger}]=\delta_{mn}$ and the same holds for all other operators. Then the relations between $n$-th
input and output operators are given by a 3x3 matrix $\mathbf{G}_{n}$
of real values equivalent to a diagram given in Fig.~\ref{fig:rw-decompos}:

\begin{equation}
\begin{pmatrix}\hat{b}_{n}^{\mathrm{(out)}}\\
\hat{r}_{n}^{(\mathrm{out})}\\
\hat{w}_{n}^{(\mathrm{out})\dagger}
\end{pmatrix}=\mathbb{\mathbf{{G}}}_{n}\begin{pmatrix}\hat{b}_{n}^{\mathrm{(in)}}\\
\hat{r}_{n}^{\mathrm{(in)}}\\
\hat{w}_{n}^{\mathrm{(in)}\dagger}
\end{pmatrix}\label{eq:g-sq-1}
\end{equation}

If $|\bar{{c}}_{W}|>|\bar{{c}}_{R}|$, the centerpiece is a squeezing
operation, as in Fig.~\ref{fig:rw-decompos}(a):

\begin{equation}
\mathbf{{G}}_{n}=(1\oplus R(\zeta))\left[\begin{pmatrix}\mathrm{{cosh}}(\xi_{n}) & \mathrm{{sinh}}(\xi_{n})\\
\mathrm{{sinh}}(\xi_{n}) & \mathrm{{cosh}}(\xi_{n})
\end{pmatrix}\oplus(-1)^n\right](1\oplus R^{-1}(\zeta))\label{eq:g-sq}
\end{equation}

When $|\bar{{c}}_{W}|<|\bar{{c}}_{R}|$, squeezing is replaced by
a beamsplitter as depicted in Fig.~\ref{fig:rw-decompos}(b):

\begin{equation}
\mathbf{G}_{n}=(1\oplus R(\zeta))\left[\begin{pmatrix}\sqrt{{1-\tau_{n}^{2}}} & \tau_{n}\\
-\tau_{n} & \sqrt{{1-\tau_{n}^{2}}}
\end{pmatrix}\oplus(-1)^n\right](1\oplus R^{-1}(\zeta))\label{eq:g-bs}
\end{equation}

\begin{figure}
\includegraphics[scale=1.2]{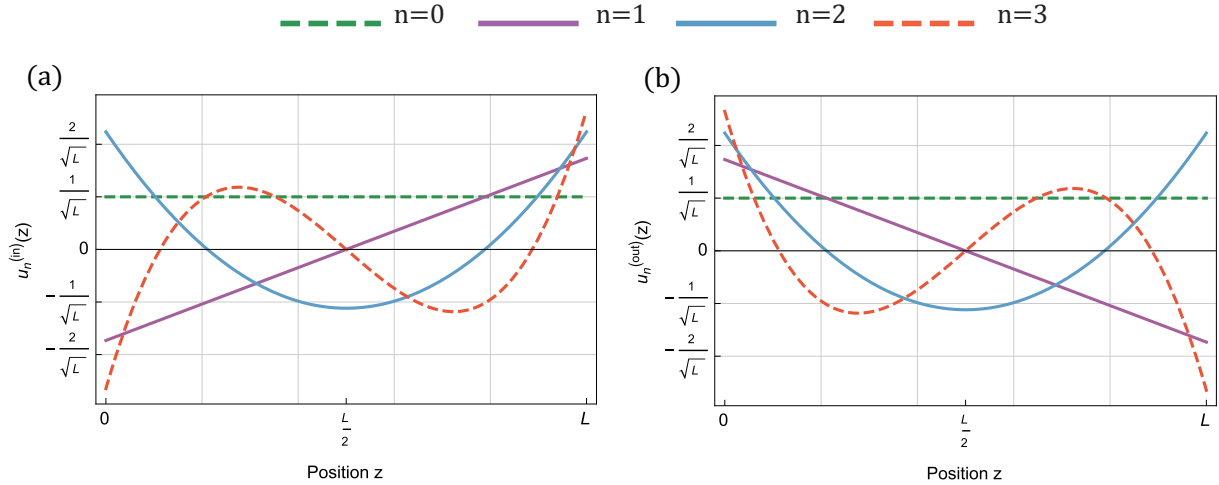}\protect\caption{(color online) Atomic (a) input $u_n^{in}(z)$ and (b) output $u_n^{out}(z)$ (b) mode functions for first four coupled modes, including the flat fundamental mode ($n=0$).  \label{fig:modes}}
\end{figure}

The mode basis for atomic $u_{n}^{({\rm in/out)}}(z)$ and photonic
fields $v_{n}^{({\rm in/out)}}(t)$ as well as squeezing
$\xi_{n}$ or beamsplitter transmission $\tau_{n}$ for the central
operation are calculated from singular value decomposition of Green
functions from Eqs.~(\ref{eq:b-sol2-1}) and (\ref{eq:c-sol2-1})
\cite{Raymer2004,Wasilewski2006,Koodynski2012}. Normalized to length
of the atomic sample $L$ and pump pulse duration $T$ they depend
solely on the product $\kappa LT=(\overline{c}_{\mathrm{R}}^{2}+|\overline{c}_{\mathrm{W}}|^{2})LT$.
For weak interaction we calculate squeezing $\xi_{n}=(\kappa LT)^{n+1/2}$
if $|\bar{{c}}_{W}|>|\bar{{c}}_{R}|$ or beamsplitter transmission
$\tau_{n}=(-\kappa LT)^{n+1/2}$ if $|\bar{{c}}_{W}|<|\bar{{c}}_{R}|$.
The mode functions are in the lowest order independent of $\kappa$:
$u_{n}^{(\mathrm{in})}(z)=\frac{{1}}{\sqrt{{L(2n+1)}}}P_{n}(2z/L-1)$,
$u_{n}^{(\mathrm{out})}(z)=\frac{{1}}{\sqrt{{L(2n+1)}}}P_{n}(1-2z/L)$,
$v_{n}^{(\mathrm{in})}(z)=\frac{{1}}{\sqrt{{T(2n+1)}}}P_{n}(2t/T-1)$,
$v_{n}^{(\mathrm{out})}(t)=\frac{{1}}{\sqrt{{T(2n+1)}}}P_{n}(1-2t/T)$
where $P_{n}(x)$ is $n$-th Legendre polynomial. We plot exemplary input and output atomic mode functions in Fig. \ref{fig:modes}.

Note that in the limiting cases of pure Stokes or pure anti-Stokes
scattering corresponding to $|\bar{c}_{W}|\gg|\bar{{c}}_{R}|$ and
$|\bar{c}_{W}|\ll|\bar{{c}}_{R}|$ we recover either two-mode squeezing
of $\hat{b}_{n}$ with $\hat{w}_{n}^{\dagger}$ or a beamsplitter
transformation between $\hat{b}_{n}$ and $\hat{r}_{n}$.

\section{Results\label{sec:Decomposition-of-three}}

\subsection{Parameters and circuit diagram}

\begin{figure}
\begin{centering}
\includegraphics[scale=1]{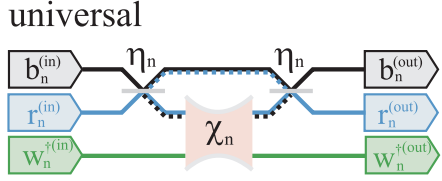}\protect\caption{(color online) Universal decomposition of Raman interaction for $n$-th
coupled triple atomic $\hat{b}_{n}$, photonic Stokes $\hat{w}_{n}$
and anti-Stokes $\hat{r}_{n}$ modes in terms of beamsplitters of
transmission $\eta_{n}$ and central squeezing by $\exp(\chi_{n})$.
Those parameters depend on both detuning from respective atomic resonances
$\Delta$ and resulting ratio of the Stokes to anti-Stokes coupling
$\bar{{c}}_{W}/\bar{{c}}_{R}$ as well as the optical depth and duration
of the interaction. Achievable range of parameters are plotted in
Fig.~\ref{fig:Transmission-of-beamsplitter}.
For simplification, in text parameters are given for $n=0$ without index. \label{fig:universal}}
\end{centering}
\end{figure}

Now we proceed with detailed interpretation of total transformation
for any mode triple. Here and in the following we drop index $n$
and consider only the most coupled mode $n=0$. In the previous section we have shown the interaction
of three coupled modes can be represented by squeezing or beamsplitter
transformation sandwiched between hyperbolic rotations as depicted
in Fig.~\ref{fig:rw-decompos}. The hyperbolic rotation can be interpreted
as another squeezing therefore we effectively describe the interaction
as a composition of squeezing operations, which makes total transformation
$\mathbf{G}$ difficult to tackle. 

To facilitate understanding and design we decompose the total transformation
$\mathbf{G}$ as a squeezing sandwiched in between beamsplitters depicted
in Fig.~\ref{fig:universal}. This form is valid for any values of
$\bar{{c}}_{W}$ and $\bar{{c}}_{R}$ and corresponds to the following
product:

\begin{equation}
\mathbf{G}=(B(\eta)\oplus1)(1\oplus S(\chi))(B(\eta)\oplus1)\label{eq:g-univ}
\end{equation}

where $B(\eta)=\begin{pmatrix}\sqrt{{1-\text{\ensuremath{\eta}}^{2}}} & \eta\\
-\eta & \sqrt{{1-\eta^{2}}}
\end{pmatrix}$ is the beamsplitter transformation between the atomic mode and anti-Stokes
mode with transmission $\eta$, and $S(\chi)=\begin{pmatrix}\mathrm{{cosh}}(\chi) & \mathrm{{sinh}}(\chi)\\
\mathrm{{sinh}}(\chi) & \mathrm{{cosh}}(\chi)
\end{pmatrix}$ is the squeezing operation.

The relation between parameters depicted in Fig.~\ref{fig:rw-decompos}
i.e. hyperbolic rotation $\zeta$ and squeezing $\xi$ or transmission
$\tau$ and parameters of general transformation is found in a closed
form. For the squeezing parameter $\chi$ we obtain:

\begin{equation}
\cosh(\chi)=\begin{cases}
\cosh^{2}(\zeta)\cosh(\xi)-\sinh^{2}(\zeta) & \mathrm{{for}\:}|\bar{{c}}_{W}|>|\bar{{c}}_{R}|\\
\cosh^{2}(\zeta)-\sqrt{{1-\tau^{2}}}\sinh^{2}(\zeta) & \mathrm{{for}\:}|\bar{{c}}_{W}|<|\bar{{c}}_{R}|
\end{cases}\label{eq:cosh-chi}
\end{equation}

and for the beamsplitter transmission $\eta$:

\begin{equation}
\eta=\begin{cases}
\sqrt{{\frac{{1+\cosh(\xi)}}{1+\cosh(\chi)}}} & \mathrm{{for}\:}|\bar{{c}}_{W}|>|\bar{{c}}_{R}|\\
\sqrt{{\frac{{1+\sqrt{{1-\tau^{2}}}}}{1+\cosh(\chi)}}} & \mathrm{{for}\:}|\bar{{c}}_{W}|<|\bar{{c}}_{R}|
\end{cases}\label{eq:eta}
\end{equation}

\begin{figure}
\begin{centering}
\includegraphics{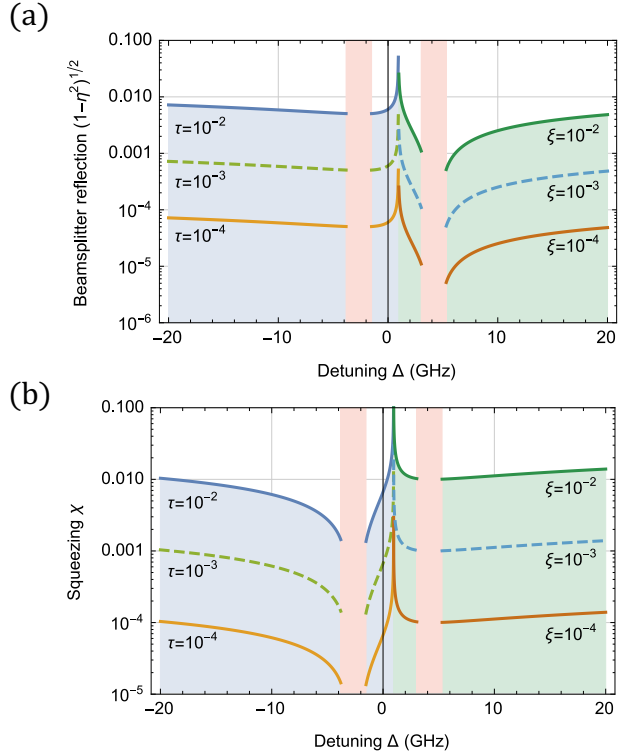}\protect\caption{(color online) (a) Transmission of beamsplitter $\eta$ and (b) squeezing
$\chi$ in the universal decomposition depicted in Fig.~\ref{fig:universal}
and described by Eq.~(\ref{eq:g-univ}) as a function of pump detuning
$\Delta$ for a various interaction parameters from decomposition
depicted in Fig.~\ref{fig:rw-decompos}: squeezing $\xi$ for domination
of the Stokes $|\bar{{c}}_{W}|>|\bar{{c}}_{R}|$ or transmission $\tau$
for domination of the anti-Stokes $|\bar{{c}}_{R}|>|\bar{{c}}_{W}|$.
\label{fig:Transmission-of-beamsplitter}}
\end{centering}
\end{figure}

In Fig.~\ref{fig:Transmission-of-beamsplitter} we plot the squeezing
$\chi$ and beamsplitter transmission $\eta$ as a function of pump detuning
$\Delta$ which determines hyperbolic rotation $\zeta$ for various
interaction strengths. We observe that not all transformations are
achievable. An ideal readout process corresponds to using 50/50 beamsplitter
i.e. $\eta=1/\sqrt{{2}}$ twice without squeezing ($\chi=0$).
At opposite extreme, spontaneous pair creation process corresponds
to transparent beamsplitters $\eta=1$ around squeezing $\chi$ in the first order equal $\sqrt{{\kappa}LT}$. We see that neither is actually possible
within the realistic model we present. To get as close as possible
to pure Stokes or anti-Stokes process we need to tune the pump close
to appropriate resonant transition, so that either $|\bar{{c}}_{W}|\gg|\bar{{c}}_{R}|$
or $|\bar{{c}}_{R}|\gg|\bar{{c}}_{W}|$.

\subsection{Output states}

In this section we study the output states of quantum memory or squeezed state generator in typical
experimental situation. First, we consider the process of spontaneous
Raman scattering, ideally realizing two-mode squeezing. The initial state of $n$-th mode is $|000\rangle_{\mathrm{{in}}}$
regardless of mode index $n$, and $|\bar{{c}}_{W}|>|\bar{c}_{R}|$.
The ket $|k,l,m\rangle_{\mathrm{{in/out}}}$ denotes $k$ excitations
in atomic spinwave mode $u_{n}^{(\mathrm{in/out})}(z)$, $l$ photons
in the anti-Stokes mode $v_{n}^{(\mathrm{in/out})}(t)$ and $m$ photons
in the Stokes mode $v_{n}^{(\mathrm{in/out})}(t)$. From the universal
decomposition depicted in Fig.~\ref{fig:universal} we can easily
obtain the output state $\mathcal{{U}}_{n}|000\rangle_{\mathrm{{in}}}$
in the Schrödinger picture with $\mathcal{U}_{n}$ being the evolution operator.
The output is most populated for $n=0$ to which we limit ourselves
dropping the index $n$ in the following consideration. The final state
is a two-mode squeezed $\exp(\chi)$ times state with one mode transmitted
trough a beamsplitter of transmission $\eta$:

\begin{equation}
\begin{aligned}
\mathcal{{U}}|000\rangle_{\mathrm{{in}}}=\frac{{1}}{\cosh(\chi)}\sum_{j=0}^{\infty}\tanh^{j}(\chi)\\
\times\sum_{k=0}^{j}\sqrt{{\binom{{j}}{k}}}(1-\eta^{2})^{k/2}\eta^{j-k}|j-k,k,j\rangle_{\mathrm{{out}}}\\
\label{eq:U0000}
\end{aligned}
\end{equation}

Note, that the above equation is valid for $|\bar{{c}}_{W}|<|\bar{c}_{R}|$
as well. As squeezing $\xi=\sqrt{(\bar{c}_{R}^{2}+|\bar{{c}}_{W}|^{2})LT}$ is small,
we may obtain the following (unnormalized) expansion:

\begin{equation}
\begin{aligned}
{\mathcal{{U}}}|000\rangle_{\mathrm{{in}}}\approx|000\rangle_{\mathrm{{out}}}+\cosh{(\zeta)}|101\rangle_{\mathrm{{out}}}\xi\\
+(\cosh^{2}(\zeta)|202\rangle_{\mathrm{{out}}}+\cosh{(\zeta)}\sinh(\zeta)|011\rangle_{\mathrm{{out}}})\xi^{2}+\ldots
\end{aligned}
\end{equation}

We observe that in the first order of interaction we indeed create
a pair of photon and an atomic excitation, which is a perfect write-in
process. In the second order we not only create two such pairs,
but also a pair of photons with each of the photons in one of the
photonic modes, corresponding to spontaneous four-wave-mixing process
with degenerate pump. Note, that the heralding of two-atom excitation
can be made by counting the photons only in the Stokes sideband. This
is possible by spectral filtering indeed necessary in many experiments
\cite{Chrapkiewicz2012,Dabrowski2014,Bashkansky2012}.

Main quantum memory operation modes are read-in, that is
mapping single photon in anti-Stokes mode to atomic mode, and the
opposite read-out operation \cite{Michelberger2015}. Transforming the operators according to the known $\mathbf{G}$ transformation,
for $\tau\ll1$ we can express the output state in the following way:

\begin{equation}
\begin{aligned}
\mathcal{{U}}|010\rangle_{\mathrm{{in}}}\\
=(-\tau\cosh(\zeta)\hat{{b}}^{\dagger}+(1-\frac{\tau^2}{2}\cosh^2(\zeta))\hat{{r}}^{\dagger}+\frac{\tau^2}{2}\cosh(\zeta)\sinh(\zeta)\hat{{w}})\mathcal{\mathcal{{U}}}|000\rangle_{\mathrm{{in}}}
\end{aligned}
\end{equation}

Where the operators $\hat{{b}}^{\dagger}$, $\hat{{r}}^{\dagger}$
and $\hat{{w}}$ are bosonic creation/annihilation operators in Schrödinger
picture acting on kets $|k,l,m\rangle$ in the usual way, and we calculate
$\mathcal{{U}}|000\rangle_{\mathrm{{in}}}$ from Eq.~(\ref{eq:U0000}). We obtain the
following expansion for the unnormalized state:

\begin{equation}
\mathcal{{U}}|010\rangle_{\mathrm{{in}}}\approx|010\rangle_\mathrm{out}+(\cosh(\zeta)|100\rangle_\mathrm{out}+\sinh(\zeta)|111\rangle_\mathrm{out})\tau\end{equation}

As expected, no state transfer occurs in the zeroth order. In the first order photon is transferred to the memory, but also a pair of Stokes photon and an atomic excitation is created. Note that in typical experimental situations $\tau$ is close to unity for many modes. In this regime contribution of Stokes photons might be even more significant.
Very similarly, when the quantum memory contains a single excitation, we may calculate how the state is transformed during read-out. For small $\tau$ the transformation is following:

\begin{equation}
\begin{aligned}
\mathcal{{U}}|010\rangle_{\mathrm{{in}}}\\
=((1-\tau^2/2)\hat{{b}}^{\dagger}+\tau\cosh(\zeta)\hat{{r}}^{\dagger}-\tau\sinh(\zeta)\hat{{w}})\mathcal{\mathcal{{U}}}|000\rangle_{\mathrm{{in}}}
\end{aligned}
\end{equation}

The output state can be expanded again, to show that in the first order pair of photons and atomic excitations are created as well:
\begin{equation}
\mathcal{{U}}|100\rangle_{\mathrm{{in}}}\approx|100\rangle_\mathrm{out}+(-\cosh(\zeta)|010\rangle_\mathrm{out}+\sqrt{\cosh(2\zeta)-1}|201\rangle_\mathrm{out})\tau\end{equation}
The above calculations show that spectral filtering is extremely important, as output state will always be polluted by spontaneously created photons even in the lowest order for weak interaction. Finally, we note that the ratio of the Stokes to anti-Stokes coupling $\bar{{c}}_{W}/\bar{{c}}_{R}$
depends solely on parameters of atoms. This ratio should be minimized
\cite{Vurgaftman2013,Zhang2014b} to ensure more faithful memory operation, as less spontaneous Stokes photons will be emitted.

\section{Conclusions\label{sec:Conclusions}}

We have introduced a model that allows us to describe Raman interaction
with both Stokes and anti-Stokes sideband present in a long atomic
sample in the presence of Doppler broadening under the weak interaction regime. As an exemplary medium
we take warm rubidium 87 vapors contained in an inert buffer gas.
The gas makes the motion of the atoms diffusive, allowing longer memory
lifetime, and induces fast velocity-class thermalization, which
we take into account in our model. Within this regime we are able
to use singular value decomposition \cite{Gerke2015,Raymer2004,Wasilewski2006,Koodynski2012}
to find modes of the interaction. In this special mode bases for the
field of atomic excitations, Stokes and anti-Stokes photons only triples
of modes are coupled, one from each of the fields. We find closed
expressions for the mode functions and coupling strengths in the weak
coupling regime.

The interaction between any triple of modes constituting Raman interface
is easiest understood when further decomposed into a product of three
two-mode quantum operations, as depicted in Fig. \ref{fig:universal}.
We give closed expressions for the parameters of constituent operations.
We analyzed which transformations are achievable in case of warm rubidium
87 vapors. 

Neither pure Stokes nor pure anti-Stokes scattering is possible, as
one process is always accompanied by another. The effects of this
interplay are easily analyzed within the framework developed. As an
example we discuss effects on the storage in and retrieval from the
quantum memory. Mapping ideal single excitation is inevitably accompanied
by creation of additional excitations due to four-wave mixing even when interaction is weak. These
processes can be suppressed by minimizing the ratio of the Stokes
to anti-Stokes coupling which is possible in atomic vapors to a limited
extent \cite{Vurgaftman2013,Zhang2014b}.

We believe that our results may serve as a zeroth-order approximation
for modeling various deleterious effects leading to decoherence such
as absorption, spontaneous emission and effects due to motion of atoms.
Multimode decoherence typically leads to quite complex results \cite{Chrapkiewicz2010}. Finally we observe
that the presented model may be applied to many similar systems, where
exact atomic structure is different\cite{Heshami2014,Parniak2015,Bustard2015,Rielander2014,Ding2015}. 

\section*{Acknowledgments}

We acknowledge insightful discussions with M. D\k{a}browski and M.
Jachura, as well as generous support of K. Banaszek. 

\section*{Funding}
This work was
supported by the National Science Center (Poland) Grant No. DEC-2011/03/D/ST2/01941
and by the Polish Ministry of Science and Higher Education \textquotedblleft Diamentowy
Grant\textquotedblright{} Project No. DI2013 011943.

\bibliographystyle{tMOP}
\bibliography{bibliography}

\begin{thebibliography}{37}
\providecommand{\natexlab}[1]{#1}
\providecommand{\noopsort}[1]{}
\providecommand{\printfirst}[2]{#1}
\providecommand{\singleletter}[1]{#1}
\providecommand{\switchargs}[2]{#2#1}

\bibitem[1]{Reim2011}
Reim, K.F.; Michelberger, P.; Lee, K.C.; Nunn, J.; Langford, N.K.; et~al. {\em
  Phys. Rev. Lett.}  {\bf 2011}, {\em 107} (5), 053603.

\bibitem[2]{Heshami2014}
Heshami, K.; Santori, C.; Khanaliloo, B.; Healey, C.; Acosta, V.M.; Barclay,
  P.E.; et~al. {\em Phys. Rev. A}  {\bf 2014}, {\em 89} (4), 040301.

\bibitem[3]{Michelberger2015}
Michelberger, P.S.; Champion, T.F.M.; Sprague, M.R.; Kaczmarek, K.T.; Barbieri,
  M.; Jin, X.M.; England, D.G.; Kolthammer, W.S.; Saunders, D.J.; Nunn, J.;
  et~al. {\em New J. Phys.}  {\bf 2015}, {\em 17} (4), 043006.

\bibitem[4]{VanderWal2003}
van~der Wal, C.H.; Eisaman, M.D.; Andr\'{e}, A.; Walsworth, R.L.; Phillips,
  D.F.; Zibrov, A.S.; et~al. {\em Science}  {\bf 2003}, {\em 301} (5630),
  196--200.

\bibitem[5]{Bashkansky2012}
Bashkansky, M.; Fatemi, F.K.; Vurgaftman, I. {\em Opt. Lett.}  {\bf 2012}, {\em
  37} (2), 142--4.

\bibitem[6]{Hosseini2009}
Hosseini, M.; Sparkes, B.M.; H\'{e}tet, G.; Longdell, J.J.; Lam, P.K.; et~al.
  {\em Nature}  {\bf 2009}, {\em 461} (7261), 241--5.

\bibitem[7]{Bimbard2014}
Bimbard, E.; Boddeda, R.; Vitrant, N.; Grankin, A.; Parigi, V.; Stanojevic, J.;
  Ourjoumtsev, A.; et~al. {\em Phys. Rev. Lett.}  {\bf 2014}, {\em 112} (3),
  033601.

\bibitem[8]{Ding2015a}
Ding, D.S.; Zhang, W.; Zhou, Z.Y.; Shi, S.; Shi, B.S.; et~al. {\em Nature
  Photon.}  {\bf 2015}, {\em 9} (5), 332--338.

\bibitem[9]{Duan2001}
Duan, L.M.; Lukin, M.D.; Cirac, J.I.; et~al. {\em Nature}  {\bf 2001}, {\em
  414} (6862), 413--8.

\bibitem[10]{Chrapkiewicz2012}
Chrapkiewicz, R.; Wasilewski, W. {\em Opt. Express}  {\bf 2012}, {\em 20} (28),
  29540.

\bibitem[11]{Koodynski2012}
Ko{\l}ody{\'n}ski, J.; Chwede{\'n}czuk, J.; Wasilewski, W. {\em Phys. Rev. A}
  {\bf 2012}, {\em 86} (1), 013818.

\bibitem[12]{Higginbottom2012}
Higginbottom, D.B.; Sparkes, B.M.; Rancic, M.; Pinel, O.; Hosseini, M.; Lam,
  P.K.; et~al. {\em Phys. Rev. A}  {\bf 2012}, {\em 86} (2), 023801.

\bibitem[13]{Dabrowski2014}
D{\k a}browski, M.; Chrapkiewicz, R.; Wasilewski, W. {\em Opt. Express}  {\bf
  2014}, {\em 22} (21), 26076.

\bibitem[14]{DeEchaniz2008}
de~Echaniz, S.R.; Koschorreck, M.; Napolitano, M.; Kubasik, M.; et~al. {\em
  Phys. Rev. A}  {\bf 2008}, {\em 77} (3), 032316.

\bibitem[15]{Brannan2014}
Brannan, T.; Qin, Z.; MacRae, A.; et~al. {\em Opt. Lett.}  {\bf 2014}, {\em 39}
  (18), 5447--5450.

\bibitem[16]{Wu2010}
Wu, C.; Raymer, M.G.; Wang, Y.Y.; et~al. {\em Phys. Rev. A}  {\bf 2010}, {\em
  82} (5), 053834.

\bibitem[17]{Wasilewski2009}
Wasilewski, W.; Fernholz, T.; Jensen, K.; Madsen, L.S.; Krauter, H.; Muschik,
  C.; et~al. {\em Opt. Express}  {\bf 2009}, {\em 17} (16), 14444.

\bibitem[18]{Sewell2013}
Sewell, R.J.; Napolitano, M.; Behbood, N.; Colangelo, G.; et~al. {\em Nature
  Photon.}  {\bf 2013}, {\em 7} (7), 517--520.

\bibitem[19]{Wasilewski2006}
Wasilewski, W.; Raymer, M.G. {\em Phys. Rev. A}  {\bf 2006}, {\em 73} (6),
  063816.

\bibitem[20]{Raymer2004}
Raymer, M.G. {\em J. Mod. Opt.}  {\bf 2004}, {\em 51} (12), 1739--1759.

\bibitem[21]{Holstein1940}
Holstein, T.; Primakoff, H. {\em Phys. Rev.}  {\bf 1940}, {\em 58} (12),
  1098--1113.

\bibitem[22]{Parniak2014}
Parniak, M.; Wasilewski, W. {\em Appl. Phys. B}  {\bf 2014}, {\em 116} (2),
  415--421.

\bibitem[23]{Chrapkiewicz2014b}
Chrapkiewicz, R.; Wasilewski, W.; Radzewicz, C. {\em Opt. Commun.}  {\bf 2014},
  {\em 317}, 1--6.

\bibitem[24]{Manz2007}
Manz, S.; Fernholz, T.; Schmiedmayer, J.; et~al. {\em Phys. Rev. A}  {\bf
  2007}, {\em 75} (4), 040101.

\bibitem[25]{McGuyer2012}
McGuyer, B.H.; Marsland~III, R.; Olsen, B.A.; et~al. {\em Phys. Rev. Lett.}
  {\bf 2012}, {\em 108} (18), 183202.

\bibitem[26]{Raymer1981}
Raymer, M.G.; Mostowski, J. {\em Phys. Rev. A}  {\bf 1981}, {\em 24} (4),
  1980--1993.

\bibitem[27]{Marsland2012}
Marsland~III, R.; McGuyer, B.H.; Olsen, B.A.; et~al. {\em Phys. Rev. A}  {\bf
  2012}, {\em 86} (2), 023404.

\bibitem[28]{Kryszewski1997}
Kryszewski, S.; Gondek, J. {\em Phys. Rev. A}  {\bf 1997}, {\em 56} (5),
  3923--3936.

\bibitem[29]{Raymer1985a}
Raymer, M.G.; Walmsley, I.A.; Mostowski, J.; et~al. {\em Phys. Rev. A}  {\bf
  1985}, {\em 32} (1), 332--344.

\bibitem[30]{Vurgaftman2013}
Vurgaftman, I.; Bashkansky, M. {\em Phys. Rev. A}  {\bf 2013}, {\em 87} (6),
  063836.

\bibitem[31]{Zhang2014b}
Zhang, K.; Guo, J.; Chen, L.Q.; Yuan, C.; Ou, Z.Y.; et~al. {\em Phys. Rev. A}
  {\bf 2014}, {\em 90} (3), 033823.

\bibitem[32]{Gerke2015}
Gerke, S.; Sperling, J.; Vogel, W.; Cai, Y.; Roslund, J.; Treps, N.; et~al.
  {\em Phys. Rev. Lett.}  {\bf 2015}, {\em 114} (5), 050501.

\bibitem[33]{Chrapkiewicz2010}
Chrapkiewicz, R.; Wasilewski, W. {\em J. Mod. Opt.}  {\bf 2010}, {\em 57} (5),
  345--355.

\bibitem[34]{Parniak2015}
Parniak, M.; Wasilewski, W. {\em Phys. Rev. A}  {\bf 2015}, {\em 91} (2),
  023418.

\bibitem[35]{Bustard2015}
Bustard, P.J.; Erskine, J.; England, D.G.; Nunn, J.; Hockett, P.; Lausten, R.;
  Spanner, M.; et~al. {\em Opt. Lett.}  {\bf 2015}, {\em 40} (6), 922--5.

\bibitem[36]{Rielander2014}
Riel{\"a}nder, D.; Kutluer, K.; Ledingham, P.M.; G{\"u}ndo{\u g}an, M.; Fekete,
  J.; Mazzera, M.; et~al. {\em Phys. Rev. Lett.}  {\bf 2014}, {\em 112} (4),
  040504.

\bibitem[37]{Ding2015}
Ding, D.S.; Jiang, Y.K.; Zhang, W.; Zhou, Z.Y.; Shi, B.S.; et~al. {\em Phys.
  Rev. Lett.}  {\bf 2015}, {\em 114} (9), 093601.

\end{thebibliography}

\end{document}